\documentclass[conference]{IEEEtran}

\IEEEoverridecommandlockouts
\usepackage{cite}
\usepackage{amsmath,amssymb,amsfonts}
\usepackage{algorithmic}
\usepackage{graphicx}
\usepackage{textcomp}
\usepackage{xcolor}
\usepackage{comment}
\RequirePackage{scrlfile}
\PreventPackageFromLoading{subfig}
\usepackage{subcaption,caption}
\usepackage{svg}
\usepackage[utf8]{inputenc}

\def\BibTeX{{\rm B\kern-.05em{\sc i\kern-.025em b}\kern-.08em
    T\kern-.1667em\lower.7ex\hbox{E}\kern-.125emX}}
    
\newcommand{\bPhi}{\boldsymbol{\Phi}}
\newcommand{\bLambda}{\boldsymbol{\Lambda}}
\DeclareMathOperator*{\argmin}{arg\,min}

\begin{document}

\title{Inferring the location of reflecting surfaces exploiting loudspeaker 
directivity} 

\author{\IEEEauthorblockN{Vincenzo 
Zaccà\IEEEauthorrefmark{1}\IEEEauthorrefmark{2} \,
Pablo Mart\'inez-Nuevo\IEEEauthorrefmark{1}\, Martin 
Møller\IEEEauthorrefmark{1}\, 
Jorge Mart\'inez\IEEEauthorrefmark{2} and Richard Heusdens\IEEEauthorrefmark{2}}

\IEEEauthorblockN{
\IEEEauthorrefmark{1}Bang \& Olufsen,
 \IEEEauthorrefmark{2}Delft University of Technology}}

\maketitle

\begin{abstract}
Accurate sound field reproduction in rooms is often limited by the lack of 
knowledge of the room characteristics. Information about the room shape or 
nearby reflecting boundaries can, in principle, be used to improve the accuracy 
of the reproduction. In this paper, we propose a method to infer the location 
of 
nearby reflecting boundaries from measurements on a microphone array. As 
opposed 
to traditional methods, we explicitly exploit the loudspeaker directivity 
model---beyond omnidirectional radiation---and the microphone array geometry. 
This approach does not require noiseless timing information of the echoes as 
input, nor a tailored loudspeaker-wall-microphone measurement step. 
Simulations show the proposed model outperforms current methods that disregard 
directivity in reberverant environments.
\end{abstract}
\section{Introduction}
The sound field produced by a loudspeaker system in an enclosed space is 
primarily---but not exclusively---determined by the loudspeakers 
characteristics 
and their position relative to the room walls \cite{Jacobsen:2013aa}. ``Smart'' 
loudspeakers including built-in microphone arrays are becoming ubiquitous. 
Often 
sound field reproduction using these high-end systems has to comply with strict 
quality requirements: The listening experience should be good irrespective of 
where the loudspeaker is placed in the room. Reflections of sound from 
the surfaces in the room --- the locations of which are unknown in practice --- 
result in inaccurate sound reproduction \cite{betlehem2005theory}. One could 
try to infer a total or partial estimate of the room shape using the built-in 
microphones, and use this information to improve the quality of the sound field 
generated by the loudspeaker system.

The problem addressed in this paper is the following: given a system composed 
of a co-located microphone array of known geometry and a co-located loudspeaker 
set with known directivity; estimate the location of reflecting surfaces close 
to the system.

Existing methods that estimate the location of reflecting surfaces by emitting 
a 
known signal using a loudspeaker are classically carried out as follows: First, 
the room impulse response (RIR) is estimated as a step in calculating the time 
of arrival (TOA). GCC-PHAT \cite{tervo2012acoustic} is an established algorithm 
to achieve this. Further consider a compact microphone array, beamforming can 
be 
used to calculate a steered-response. This results in improved robustness 
against uniform spatially uncorrelated noise. In this setting, the microphone 
array geometry can be exploited as prior for the TOA estimation.     
In a scenario where the echoes need to be sorted, greedy methods are often used 
\cite{coutino2017greedy,iJager2016}. The performance of these methods 
usually degrades with reverberation---i.e. reflecting boundaries---, especially 
when received echoes have overlap in time (which happens due to 
finite measurement bandwidth). Moreover, the echo sorting problem is 
computationally demanding.

The problem can be relaxed by including information about the loudspeaker-wall 
behavior and solving the source localization problem jointly. In particular, 
explicit loudspeaker modeling is often neglected and simplified to 
omnidirectional or highly directive models \cite{antonacci2012inference}. In 
\cite{ribeiro2011geometrically}, the loudspeaker-wall interaction is implicitly 
considered by constructing a dictionary from experimental measurements. That 
method improves performance in an ideal scenario, but it is not robust when the 
scenario deviates from the measured dictionary. 

In this paper, we propose a measurement model that explicitly includes 
loudspeaker directivity and the microphone array geometry. We use this model to 
solve an inverse problem: from microphone measurements it outputs an estimation 
of the nearby reflecting boundaries. This approach does not require noiseless 
timing information of reflections, nor a tailored loudspeaker-wall-microphone 
measurement dictionary. As indicated by the simulations, the proposed algorithm 
shows improved performance compared to current methods that disregard 
loudspeaker directivity.
\section{Proposed method}
We propose a novel microphone signal model that maps the location of image 
sources to microphone measurements. Then, the problem of estimating the 
location 
of reflecting surfaces is solved as an inverse problem. We first describe the 
signal model in the continuous domain. After discretization it is reformulated 
in matrix-vector form. Finally, the inverse problem is posed as an 
$\ell_1$-regularized convex optimization problem.

Similar to the image source method \cite{allen1979image}, our signal model is 
based on geometric acoustics, i.e. the concept of sound waves is replaced by 
sound rays that travel in a narrow path and reflect specularly. We assume that 
all image sources lie on a horizontal plane in order to model vertical planar 
surfaces in the room. Although presumably extensible to three dimensions, we 
consider for simplicity that the ceiling and floor reflectors fully absorb. We 
further assume that the center of the loudspeaker system lies at the origin, 
thus coinciding with the geometric center of a uniform circular array.
  \begin{figure}[t]
  \centering
	\def\svgwidth{0.8\linewidth}
%% Accompanies image file 'geometryangles.pdf' (pdf, eps, ps)
\begingroup%
  \makeatletter%
  \providecommand\color[2][]{%
    \errmessage{(Inkscape) Color is used for the text in Inkscape, but the package 'color.sty' is not loaded}%
    \renewcommand\color[2][]{}%
  }%
  \providecommand\transparent[1]{%
    \errmessage{(Inkscape) Transparency is used (non-zero) for the text in Inkscape, but the package 'transparent.sty' is not loaded}%
    \renewcommand\transparent[1]{}%
  }%
  \providecommand\rotatebox[2]{#2}%
  \newcommand*\fsize{\dimexpr\f@size pt\relax}%
  \newcommand*\lineheight[1]{\fontsize{\fsize}{#1\fsize}\selectfont}%
  \ifx\svgwidth\undefined%
    \setlength{\unitlength}{2700bp}%
    \ifx\svgscale\undefined%
      \relax%
    \else%
      \setlength{\unitlength}{\unitlength * \real{\svgscale}}%
    \fi%
  \else%
    \setlength{\unitlength}{\svgwidth}%
  \fi%
  \global\let\svgwidth\undefined%
  \global\let\svgscale\undefined%
  \makeatother%
  \begin{picture}(1,0.49185185)%
    \lineheight{1}%
    \setlength\tabcolsep{0pt}%
    \put(0,0){\includegraphics[width=\unitlength,page=1]{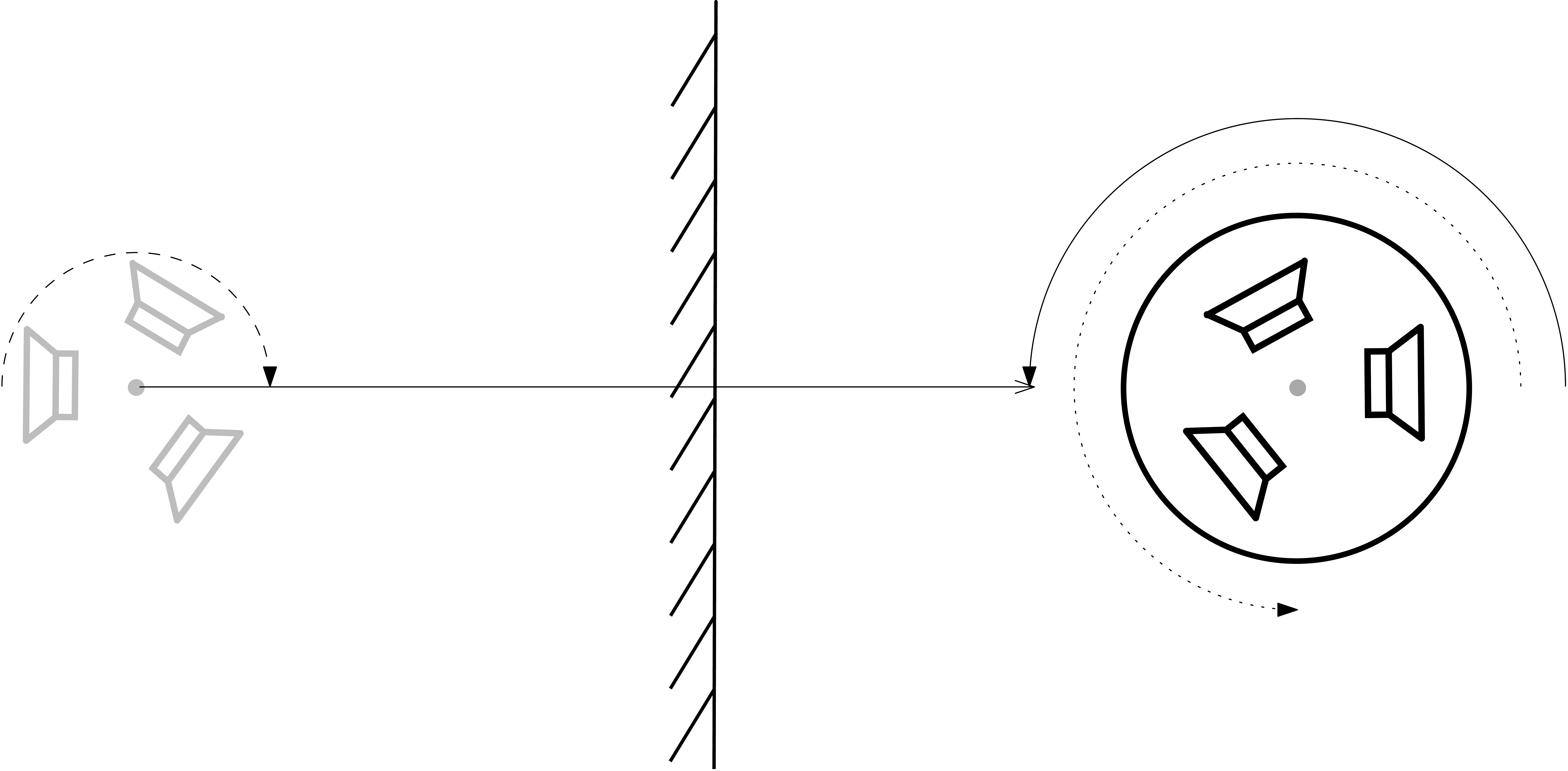}}%
    \put(0.17860282,0.2585617){\color[rgb]{0,0,0}\makebox(0,0)[lt]{\lineheight{1.25}\smash{\begin{tabular}[t]{l}$\varphi$\end{tabular}}}}%
    \put(0.61021054,0.2565694){\color[rgb]{0,0,0}\makebox(0,0)[lt]{\lineheight{1.25}\smash{\begin{tabular}[t]{l}$\phi$\end{tabular}}}}%
    \put(0.81238841,0.06292151){\color[rgb]{0,0,0}\makebox(0,0)[lt]{\lineheight{1.25}\smash{\begin{tabular}[t]{l}$\theta$\end{tabular}}}}%
    \put(0.00805795,0.44739974){\color[rgb]{0,0,0}\makebox(0,0)[lt]{\lineheight{1.25}\smash{\begin{tabular}[t]{l}Image source\end{tabular}}}}%
    \put(0.66799754,0.44454351){\color[rgb]{0,0,0}\makebox(0,0)[lt]{\lineheight{1.25}\smash{\begin{tabular}[t]{l}Loudspeaker system\end{tabular}}}}%
    \put(0,0){\includegraphics[width=\unitlength,page=2]{figures/geometryangles.pdf}}%
    \put(0.46308905,0.4440207){\color[rgb]{0,0,0}\makebox(0,0)[lt]{\lineheight{1.25}\smash{\begin{tabular}[t]{l}Wall\end{tabular}}}}%
  \end{picture}%
\endgroup%

	\caption{Model setup. To the right a loudspeaker system and a circular 
microphone array. The center dot represents the origin of coordinates. To the 
left an image source that models sound reflection. A reflected sound ray is 
depicted. The angles $\phi$, $\varphi$, and $\theta$ denote the angle of the 
image source position, the emission angle, and the angle representing a 
position on the microphone array, respectively.
}
	\label{fig:geometry}
\vspace{-0.4cm}
\end{figure}
\subsection{Continuous Measurement Model}
Our model is determined by three main components: Specular reflection modeled 
by equivalent image sources, a loudspeaker with a non-homogeneous directivity 
response, and a co-located circular microphone array. 
Consider the set $\mathcal{S}$ containing the spatial locations of $K$ image 
sources of first order, i.e. $\mathcal{S} = \{\mathbf{r}_k\}^{K-1}_{k=0}$ for 
$\mathbf{r}_k\in\mathbb{R}^2$ and $K>0$. In this work we restrict ourselves to 
first-order images since we are considering only dominant reflections. In 
general the contributions of these sources dominate the early part of the room 
impulse response. For convenience spatial locations are expressed in polar 
coordinates, i.e. $\mathbf r=(R,\phi)$, for $R\in \mathbb{R}_{+}$ and $\phi \in 
\mathcal{E} = [0,2\pi)$ the angular support. 

We start the analysis describing a theoretical circular array with a continuous 
aperture. A practical microphone array is later modeled by sampling the 
continuous aperture in space. Let the signal received by the continuous array 
be 
denoted by $y(t,\theta)$, with $\theta\in \mathcal{E}$. Here we assume a fixed 
array radius. Dependence on the radius is therefore not stated explicitly. The 
signal model is given by the following map:
\vspace{-0.1cm}
\begin{IEEEeqnarray}{rCl}
\label{eq:CTsignalmodel}
\ &\mathbb{R}^2\ &\to\ L^2(\mathbb{R},\mathcal{E})\\
&\mathcal{S}\ &\mapsto\ y(t,\theta),\nonumber
\vspace{-0.2cm}
\end{IEEEeqnarray}
where $L^2$ is Lebesgue's space of finite energy signals. In this paper, the  
goal is to obtain an approximation of the inverse of the map stated in
(\ref{eq:CTsignalmodel}), i.e. to recover the image source locations from 
signal 
measurements at different angles (i.e. array positions). In later sections we 
show that the inverse problem can be posed as a convex optimization problem in 
the discrete domain. The three main components of our model are introduced next.
\subsubsection{Image Source locations} The loudspeaker system emits a known 
signal $x(t)$. The sound is reflected specularly at a wall. Estimation of the 
source location is derived from an estimation of the wall's position. This is 
depicted in Fig. \ref{fig:geometry}. The received signal $y$ can be written as
\begin{equation}
\label{eq:CTRXsignal}
    	y(t,\theta) = \left(x\ast\left(h_{\text{dp}}(\cdot,\theta) + 
\sum_{\mathbf{r}\in \mathcal{S}} 
h_s(\cdot,\theta,\mathbf{r})\right)\right)\big(t\big),
\end{equation}
where $\ast$ denotes convolution, $h_{\text{dp}}(t,\theta)$ is the channel 
response of the direct path from the loudspeaker to a microphone located at 
$\theta$, and $h_s(t,\theta,\mathbf{r})$ models the path from an image source 
to 
a point on the circular array. The direction of arrival of sound rays 
corresponding to first (and in some cases second) order reflections coincides 
with the angle formed by the wall's normal point and the center of the 
array, as seen in Fig. \ref{fig:geometry}.
\subsubsection{Loudspeaker Directivity}
Let us define an angle-dependent loudspeaker system response in the far-field 
(at a distance $R_0$), and denote it by $\gamma_0(t,\varphi)$, where $\varphi 
\in \mathcal{E}$. We assume this function is known (e.g. it has been measured a 
priori), and the system is linear time-invariant. Then we can extrapolate the 
loudspeaker impulse response at distances $R>0$,
\begin{equation}
\label{eq:CTdirectivity}
\frac{R_0}{R}\gamma_0\left(t-\frac{R-R_0}{v_c}, \varphi\right),
\end{equation} 
where $v_c$ denotes the speed of sound in m/s which is assumed constant and 
known. Note that (\ref{eq:CTdirectivity}) constitutes a (simplified) model for 
loudspeaker system directivity.
\subsubsection{Circular-array response}
Without loss of generality we set the center of the circular array at the 
origin of coordinates. Image sources are assumed to be in the far field. It is 
therefore reasonable to assume signal attenuation due to the distance is 
approximately equal at all positions on the array's circumference. The array 
response is then determined by relative delays between positions on the 
circumference. These delays can be inferred from the direction of arrival of 
incoming sound, and the array geometry. For the uniform circular array, it is 
easy to show, referring to (\ref{eq:CTRXsignal}) and (\ref{eq:CTdirectivity}), 
that \cite{torres2013room,brooks2006deconvolution}
\begin{equation}\label{eq:oneresponse}
h_s(t,\theta,R,\phi)\! =\! \frac{R_0}{R}\gamma_0\!\left(\! t\!-\! 
\frac{R\!-\!R_0}{v_c}\!-\!\frac{R_a}{v_c}\cos \left(\theta-\phi \right), 
\phi\!\right),
\end{equation}
where $R_a$ is the radius of the array. For this particular geometry and image 
source set $\mathcal{S}$, the angle $\varphi = \phi$ in Eq. 
(\ref{eq:oneresponse}). The function $h_s(t,\theta,R,\phi)$ represents the 
response (including loudspeaker directivity) at an angle $\theta$ in the array 
due to a image source located at $\mathbf{r}$ (recall 
that $\mathbf{r}= (R,\phi)$). In an enclosed space (e.g, a 
room), the channel response including all sources contributions is given by 
$h(t,\theta):=\sum_{\mathbf{r}\in \mathcal{S}} h_s(t,\theta,R,\phi)$. Let us 
now express $h$ as a convolution integral, this is
\begin{equation}
\label{eq:sparseconvolution}
 h(t, \theta) = \int_{\mathcal{E}}\int_{\mathbb{R_{+}}}h_s(t,\theta, 
R',\phi')u_{\mathcal{S}}(R',\phi') \mathrm{d}R' \mathrm{d}\phi',
\end{equation}
where
$u_{\mathcal{S}}(R,\phi):=\sum_{\mathbf{r'}\in\mathcal{S}}(R_0/R)\delta(R-R',
\phi-\phi')$. The key observation is that the above can be decomposed as three 
convolutions:
\begin{equation}
\begin{split}
h(t,\theta) = & \int_{\mathbb{R}}\int_{\mathcal{E}}  
\delta\left(t-t'-\frac{R_a}{v_c}\cos( \theta-\phi')\right)\\
    &\! \! \!\int_{\mathbb{R_{+}}}  \gamma_0(t' 
-\frac{R'-R_0}{v_c},\phi')u_{\mathcal{S}}(R',\phi') \,\mathrm d R' \mathrm d 
\phi'   \mathrm d t',
\end{split}
\label{eq:continuous}
\vspace{-0.2cm}
\end{equation}
which can be interpreted as a linear convolution in time---proportional to 
distance---to extrapolate the loudspeaker impulse response, and a 
two-dimensional convolution to compute the relative microphone delays. In the 
next section, we discretize (\ref{eq:continuous}) and reformulate it in 
matrix-vector form.
\subsection{Discrete Measurement Model}
The system defined by $h(t,\theta)$ is described by the model's three main 
components: the image source locations, the directivity information of the 
loudspeaker, and the influence of the microphone array geometry. It is further 
assumed that we can know a reasonably accurate discrete version of 
$h(t,\theta)$, the array geometry, and the loudspeaker's directivity. We then 
aim at exploiting this information in order to find the image source locations 
corresponding to the dominant reflecting boundaries. Note first that in a real 
scenario it is necessary to have a preprocessing stage. In particular, with the 
right choice of excitation signals it is possible to perform an accurate 
deconvolution \cite{stan2002comparison}. Moreover the direct path 
$h_{\textrm{dp}}(t,\theta)$, can in principle be estimated from anechoic 
measurements and be subtracted from the process. 

We consider that the input to our system is a discrete version of $h(t,\theta)$ 
denoted by $h[n,m]$ for time steps $n=0,\dotsc,N_h-1$ and microphone indexes 
$m=0,\dotsc,M-1$. In other words, we assume we can obtain $h[n,m]$ from 
measurements $y(t,\theta)$. We show below how to decompose this system into its 
three different building blocks.
\subsubsection{Image Source Locations}
We first discretize $\mathbb{R}^2$ uniformly in polar coordinates. Our 
microphone measurements are sampled in time at $f_s$~Hz for $M$ distinct 
microphones. We use a stepsize for the radial distance of $\Delta R = v_c/f_s$, 
and an angular stepsize $\Delta \theta = 2\pi/(MP)$ for some integer $P\geq1$ 
representing an upsampling factor in the angle domain. We restrict the image 
source distances to a range between $R_{\text{min}} = R_a$ and $R_{\text{max}} 
= 
T v_c/f_s+R_a$ for some integer $T\geq1$. Image source locations are then 
assigned to a closest point in this discretized set. We create the 
corresponding 
Voronoi regions in $\mathbb{R}^2$ by using this discrete set of points as 
generators. Thus, we have $TMP$ Voronoi regions denoted as $V(\mathbf{g})$ for 
generator $\mathbf{g} \in \mathbb{R}^2$. We define a two-dimensional function 
that conveys the information about the sources locations, i.e.
\vspace{-0.1cm}%
\begin{equation}
s[q,p]:=\sum_{k=0}^{K-1}\frac{R_0}{R_k}\mathbf{1}_{V(q\frac{v_c}{fs},p\frac{
2\pi}{MP})}(\mathbf{r}_k),
\end{equation}
\vspace{-0.1cm}
where the generator is given in polar coordinates and the ranges of $q$ and $p$ 
follow from the definitions above. The indicator function 
$\mathbf{1}_{V(q\frac{v_c}{fs},p\frac{2\pi}{MP})}(\mathbf{r}_k)$, takes the 
value 1 whenever the $k$th image source location falls in the $q$th, $p$th 
Voronoi region, and $0$ otherwise. In other words, $s[q,p]$ represents the 
different modeled spatial locations. 
\subsubsection{Loudspeaker Directivity}
The loudspeaker model $\gamma_0(t,\theta)$ representing the directivity of the 
loudspeaker is discretized as 
\begin{equation}
\label{eq:angleIR}
v[n,p]:=\gamma_0\Big(\frac{n}{f_s} -\frac{R_0f_s}{v_c},\frac{2\pi p}{MP}\Big),
\end{equation}
for $n=0,\ldots,N_v-1$ and $p=0,\ldots,MP-1$. 
\subsubsection{Array Geometry} 
Microphone signals are obtained by spatial sampling of the model's continuous 
aperture. The microphone positions on the circular aperture are modeled as,
\begin{equation}
\mu[n,p]:= \begin{cases}
1 & \text{if } n = \left \lceil f_s \frac{R_a}{v_c}\left(1- 
\cos\left(\frac{2\pi p}{MP}\right)\right) \right \rceil, \\
0 & \text{otherwise}
\end{cases},
\end{equation}
for $n=0,\ldots,N_\mu-1$, where $\lceil \cdot \rceil$ denotes the ceiling 
operator. Analogous to (\ref{eq:continuous}), we express $h[n,m]$ in terms of 
two-dimensional and one-dimensional discrete convolutions, i.e.
\begin{equation}
\label{eq:Convs}
\begin{split}
h[n,m] = & \sum_{m'=0}^{M-1} \sum_{n_h=0}^{N_h-1} \mu\left [n-n_h,\left(mP-m' 
\right )_{\text{mod }MP}\right]\\
&\quad \quad \ \sum_{n_v=0}^{N_v-1} v[n_h-n_v,m']s[n_v,m']\\ 
= & \mu[nP,m] \ast_{n,m}\big(v[n, m]\ast_n s[n,m] \big ).
\end{split}
\end{equation}
\subsection{Inverse Problem}
We pose the inverse problem as a linear system of equations. In this manner, we 
relate the vector of image locations to the linear system estimates at each of 
the microphones. Let us define $\mathbf{s}^{(p)}$ as a vector of size $T$, 
with elements $\mathbf{s}_q^{(p)}:=s[q,p]$, and let
\vspace{-0.2cm}
\begin{equation}
\mathbf{s}:=\left[[\mathbf{s}^{(0)}]^{\top},\ldots,[\mathbf{s}^{(MP-1)}]^{
\top}\right ]^{\top},
\end{equation}
where $^\top$ denotes transposition. Note $\mathbf{s}$ is of size $TMP$. The 
channel impulse responses to each microphone are arranged in a vector of size 
$N_hM$ as
\begin{equation}\label{eq:defhcolvec}
\mathbf{h} :=\!\left[ [\mathbf{h}^{(0)}]^{\top}\!,\ldots,[ 
\mathbf{h}^{(m)}]^{\top}\!,\ldots,[\mathbf{h}^{(M-1)}]^{\top}\right]^{\top},
\end{equation} 
where $\mathbf{h}^{(m)}$ has size $N_h$, with elements 
$\mathbf{h}_n^{(m)}:=h[n,m]$; it is the vector representation of the impulse 
response to microphone $m$. The forward model is now posed as
\begin{equation}
    \mathbf{h} = \bPhi\mathbf{s} + \mathbf{n},
    \label{eqn:forwardmodel}
\end{equation}
where $\bPhi$ is a matrix of size $N_hM\times TMP$ representing the operation 
in 
Eq.~(\ref{eq:Convs}), and $\mathbf{n}$ is a noise term. 

We conclude this section by showing how matrix $\bPhi$ is explicitly 
constructed. In brief we show that it has a block Toeplitz structure which, 
after proper zero-padding makes it amenable to implement using the FFT. We 
model the microphone array and the loudspeaker directivity contributions using 
matrices $\mathbf{A}$ and $\mathbf{D}$, respectively. Let $\mathbf{I}_N$ denote 
the identity matrix of size $N \times N$ and let us define the zero-padding 
matrix $\mathbf{W}_{a\times b}$ as
\begin{equation}
    \mathbf{W}_{a\times b} = \begin{bmatrix}
\mathbf{I}_{b}\\ \mathbf{0}_{a-b \times b}
\end{bmatrix},
\end{equation}
for some positive integers $a\geq b$. Denote by $\mathbf{F}_M$ the normalized 
DFT matrix of size $M \times M$. Then,
\begin{equation}\label{eq:phi}
    \bPhi := \left(\mathbf{I}_{MN_h} \otimes \begin{bmatrix}
    1, \mathbf{0}_{P-1}
    \end{bmatrix} \right)\mathbf{A}\mathbf{D} \left( \mathbf{I}_{MP}\otimes 
\mathbf{W}_{N_h \times T}\right).
\end{equation}
where $\otimes$ is the Kronecker product. We give next explicit expressions for 
the matrices $\mathbf{A}$ and $\mathbf{D}$.
\subsubsection{Loudspeaker Directivity $\mathbf{D}$}
We form a vector by concatenating the angle-dependent directivity 
responses in (\ref{eq:angleIR}) 
\begin{equation}
\mathbf{v} \!:=\! 
\left[[\mathbf{v}^{(0)}]^{\top}\!,\ldots,[\mathbf{v}^{(p)}]^{\top}\!,\ldots 
[\mathbf{v}^{(MP-1)}]^{\top}\right]^{\top},
\end{equation}
of size $N_vMP$, where $\mathbf{v}^{(p)}$ has size $N_v$ and elements 
$\mathbf{v}_{n}^{(p)}:=v[n,p]$. Then we define
\begin{equation}
\mathbf{D}:=\left (\mathbf{I}_{MP} \otimes \mathbf{F}_{N_h}\right 
)^{-1}\bLambda_{\mathbf{v}}\left (\mathbf{I}_{MP} \otimes 
\mathbf{F}_{N_h}\right 
),
\end{equation}
where
\vspace{-0.2cm}
\begin{equation}
\bLambda_{\mathbf{v}}:=\text{diag} \left\{ \left(\mathbf{I}_{MP} \otimes 
\mathbf{F}_{N_h} \mathbf{W}_{(N_h \times N_v)} \right)
\mathbf{v}
\right\},
\end{equation}
where $\mathrm{diag}\{\mathbf{b}\}$ is a diagonal matrix with entries given by 
the elements in vector $\mathbf{b}$.
\subsubsection{Array Geometry $\mathbf{A}$} We make a vector of size $N_\mu MP$
\begin{equation}
\mathbf{m}\!:=\!\!\left[[\mathbf{m}^{(0)}]^{\top}\!,\ldots,\![\mathbf{m}^{(MP-1)
}]^{\top}\right]^{\top},
\end{equation}
where $\mathbf{m}^{(p)}$ has size $N_{\mu}$, and has elements 
$\mathbf{m}_{n}^{(p)}:=\mu[n,p]$. Then,
\begin{equation}
\mathbf{A} := \left(\mathbf{F}_{MP} \otimes \mathbf{F}_{N_h} \right)^{-1} 
\bLambda_{\mathbf{m}} \left(\mathbf{F}_{MP} \otimes \mathbf{F}_{N_h} \right), 
\end{equation}
where
\begin{equation}
\bLambda_{\mathbf{m}}\!:=\! \text{diag}\left\{(\mathbf{F}_{MP}\! \otimes 
\mathbf{F}_{N_h}) \left(\mathbf{I}_{MP} \otimes \mathbf{W}_{(N_h \times 
N_\mu)}\right) \mathbf{m} \right\}.
\end{equation}
%%%%%%%%%%%%%
\subsection{Solution to the Inverse Problem}
The solution of the inverse problem consists of extracting the image source 
locations $\mathbf{s}$ from the channel estimates $\mathbf{h}$, i.e. 
an estimation of the inverse of the forward model in 
Eq.~(\ref{eqn:forwardmodel}). The system of equations in 
Eq.~(\ref{eqn:forwardmodel}) 
is in general overdetermined since $N_hM>TMP$. It is possible to interpret 
$\bPhi$ 
as a large dictionary of reflections where each column captures the channel 
response
due to a single image source. In principle, the indexes in $\mathbf{s}$ 
corresponding
to non-zero values in the solution carry image source location information. We
estimate the inverse of the forward model in Eq. (\ref{eqn:forwardmodel}) using
two different approaches.

First note that the magnitude of delay-and-sum steered-response can 
be computed using $\mathbf{A}^{\top}$. Also the matched filter for the 
loudspeaker directivity response involves multiplication with 
$\mathbf{D}^\top$. Then, the cross-correlation delay-and-sum estimate is given 
by
\vspace{-0.1cm}
\begin{equation}
\hat{\mathbf{s}}_{\text{CC-DAS}} = |\bPhi^\top\mathbf{h}|,
\end{equation}
where $|\cdot|$ denotes element-wise absolute value. It is important to 
emphasize that this is a generalization of the methods presented in 
\cite{tervo2012acoustic} and \cite{torres2013room}. Therein, the loudspeaker is 
assumed omnidirectional. These approaches are known to be biased for 
loudspeaker impulse response mismatches and for closely spaced sources 
\cite{brooks2006deconvolution}.

Moreover, note that the vector $\mathbf{s}$ is sparse since only a few 
candidate 
positions will be occupied by the unknown image sources. We therefore pursue a 
sparcity promoting solution, i.e. we write the problem as an 
$\ell_1$-regularized least squares problem,
\vspace{-0.1cm}
\begin{equation}
\hat{\mathbf{s}}_{\text{sparse}} = \argmin_{\mathbf{s}\in\mathbb{R}^{TMP}} 
\left 
\|\mathbf{h} - \bPhi\mathbf{s} \right \|_2^2 + \lambda  
\left\|\mathbf{s}\right\|_1,
\label{eq:inverse_solution}
\end{equation}
\vspace{-0.1cm}
where $\lambda>0$ is a regularization parameter.
\begin{figure}
\centering
\includegraphics[width=0.38\textwidth]{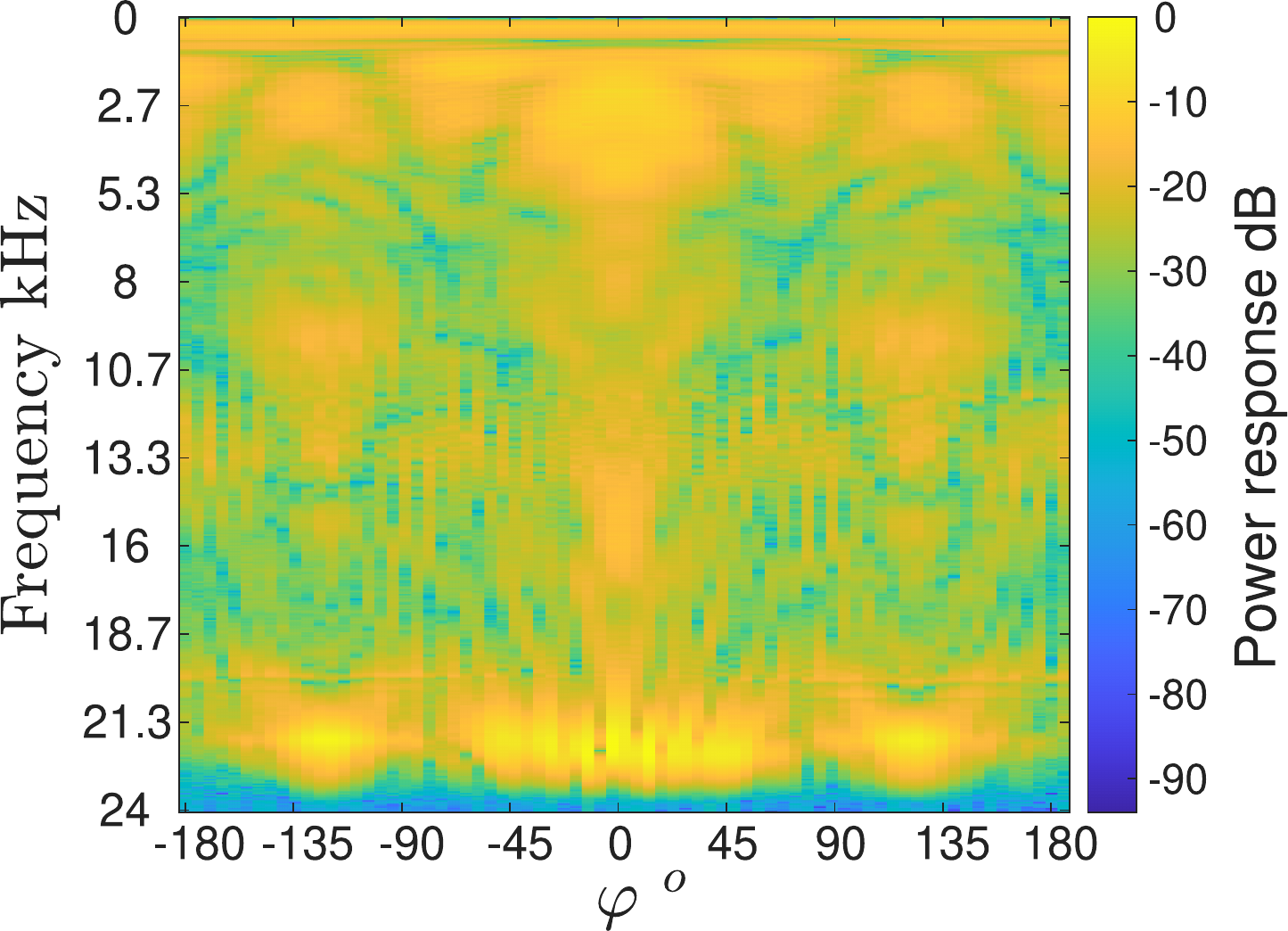}
 \caption{Measured loudspeaker directivity function.}
\label{fig:directivity_plot}
\vspace{-0.4cm}
\end{figure}
\begin{figure*}[t]
	\centering		
	\begin{subfigure}[b]{0.24\textwidth}
		\centering
		\includegraphics[width=\textwidth]{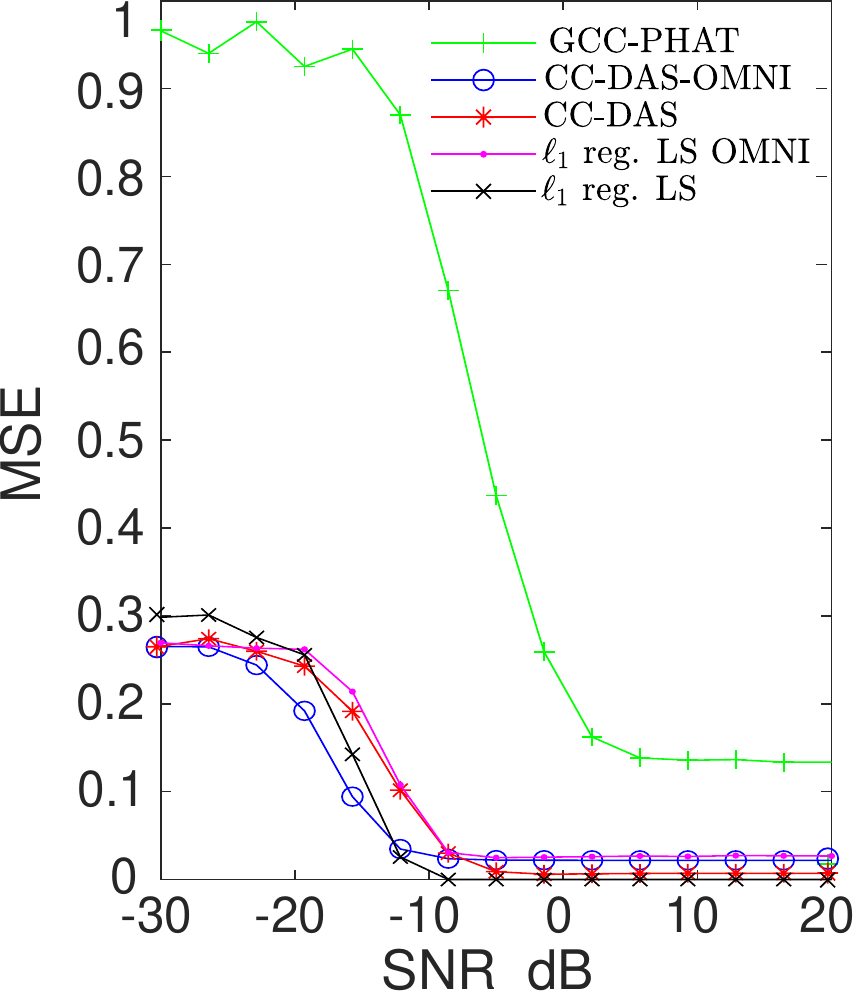}
		\caption{Single-wall MSE}
		\label{fig:res1a}
	\end{subfigure}
	\begin{subfigure}[b]{0.24\textwidth}
		\centering
    \includegraphics[width=\textwidth]{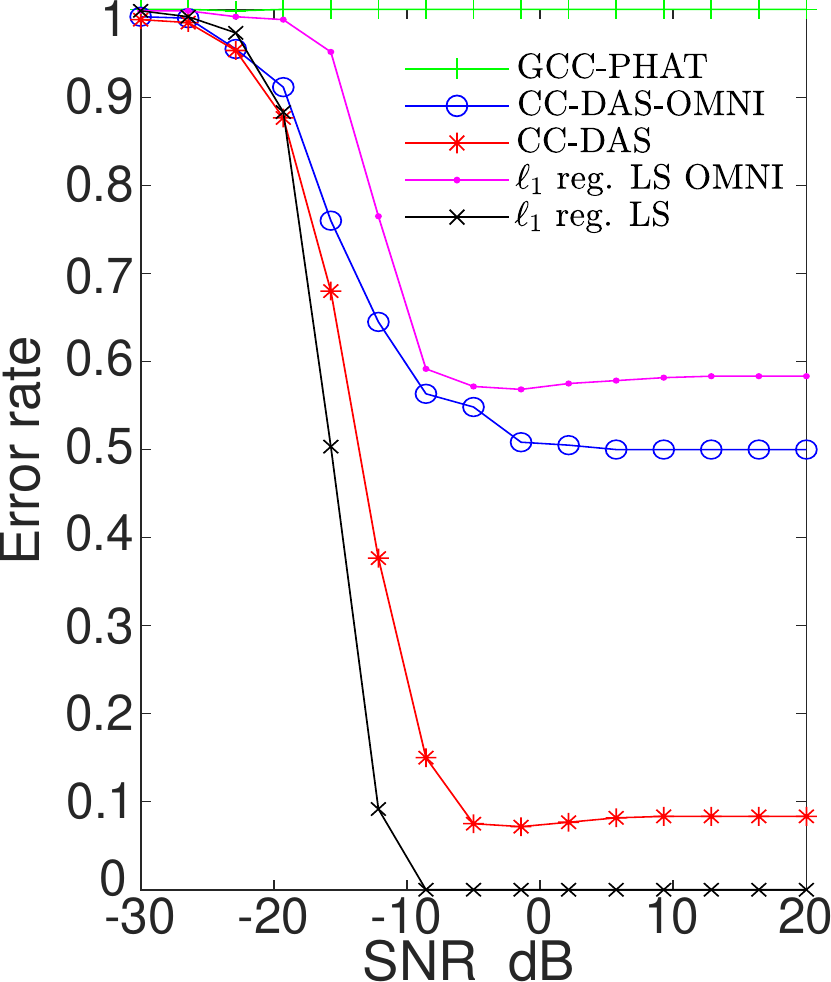}
		\caption{Single-wall error rate}
		\label{fig:res1b}
	\end{subfigure}	
	\begin{subfigure}[b]{0.24\textwidth}
		\centering
        \includegraphics[width=\textwidth]{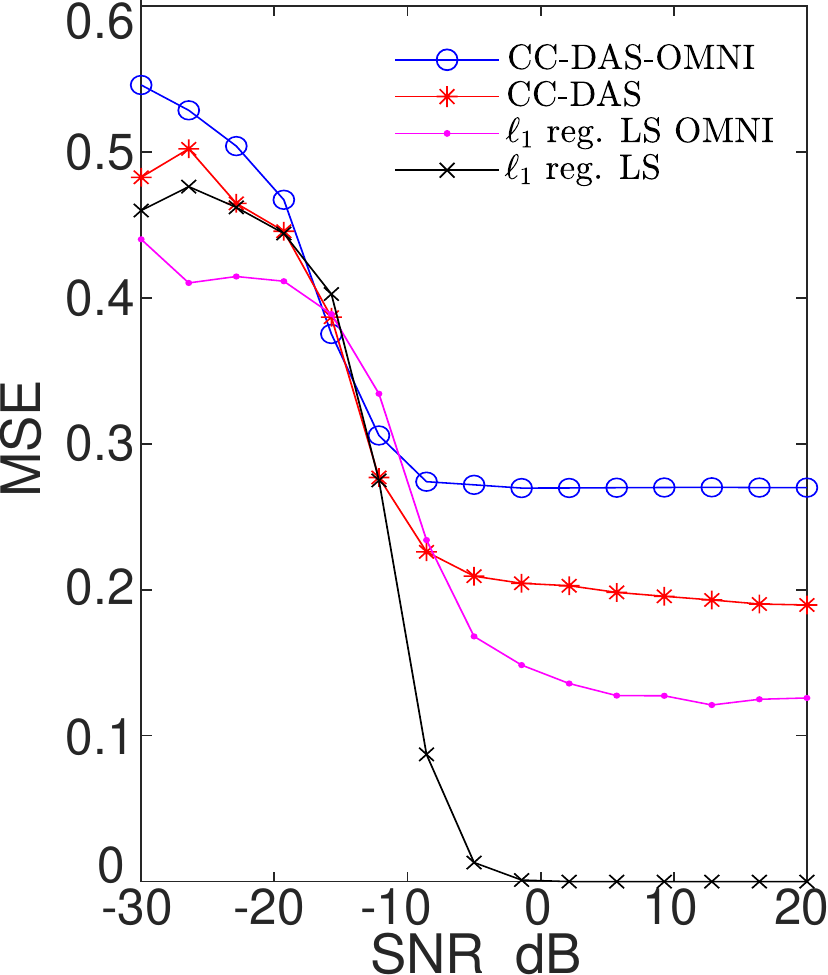}
		\caption{Corner MSE}
		\label{fig:res1c}
	\end{subfigure}
	\begin{subfigure}[b]{0.24\textwidth}
		\centering
        \includegraphics[width=\textwidth]{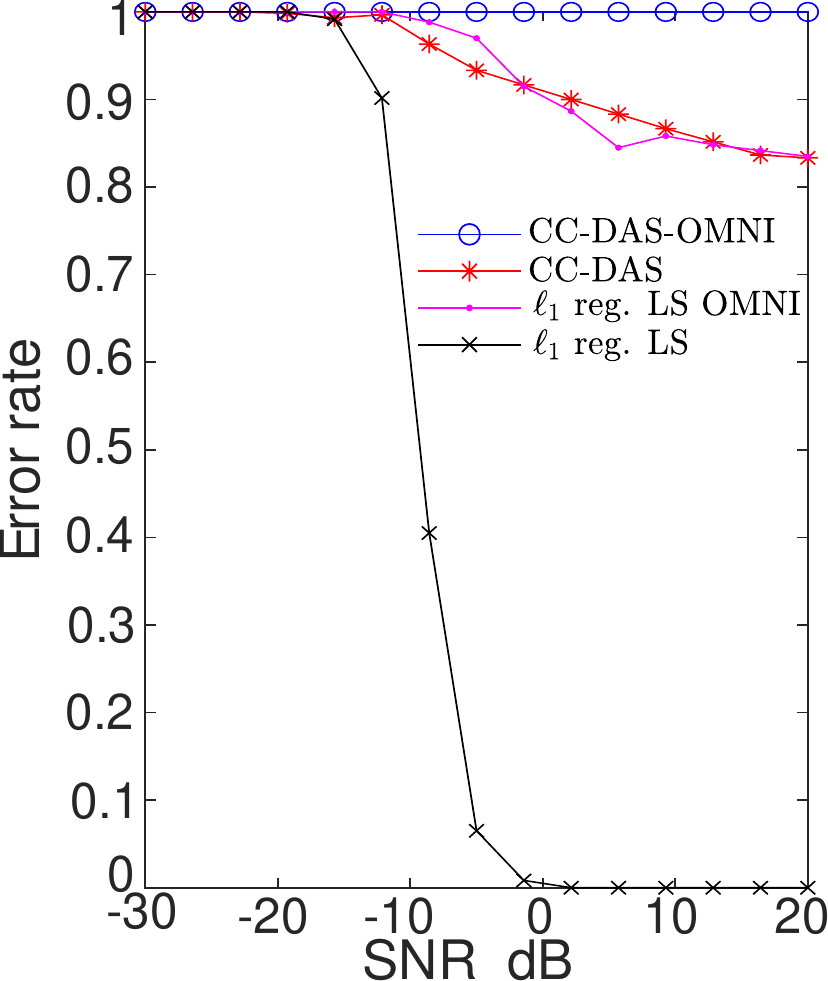}
		\caption{Corner error rate}
		\label{fig:res1d}
	\end{subfigure}	
	\caption{Performance comparison of various methods for localizing image
sources in the presence of AWGN.}
\vspace{-0.45cm}
\end{figure*}	
\section{Experimental Results}
\subsection{Setup}
We evaluate the performance of the proposed algorithm in two different 
scenarios 
and compare it with other methods. We focus on two factors that influence 
performance. First, the assumption of an omnidirectional directivity pattern 
whereas the loudspeaker is directive. Second, the ability to resolve echoes 
whenever there exist more than one reflecting boundary. In particular, we 
consider the loudspeaker deconvolution techniques GCC-PHAT 
\cite{knapp1976generalized}, cross-correlation delay-and-sum (CC-DAS) 
\cite{torres2013room,tervo2012acoustic}, and $\ell_1$-regularized least squares 
(LS)\cite{ribeiro2011geometrically}. In GCC-PHAT localization is 
performed through trilateration based on the TOA estimates of the array. An 
overview of these methods is provided in Table \ref{tab:methods}.
In high quality systems, a loudspeaker with high bandwidth that radiates 
homogeneously in a wide angle range is desired. The directivity function used 
in the experiments in depicted in Fig.~\ref{fig:directivity_plot}. It 
is obtained from measurements in anechoic conditions.
In order to assess performance in estimating location, i.e. $\hat{\mathbf{r}}$,
under additive white Gaussian noise (AWGN), we have used two different metrics: 
the mean squared error $\text{MSE}(\hat{\mathbf{r}}) =\left 
\|\mathbf{r}-\hat{\mathbf{r}} \right \|^2$, and the error rate, i.e. 
localization is incorrect if $||\mathbf{r}-\hat{\mathbf{r}}||^2>\epsilon$ for 
$\epsilon=0.01$, where $\mathbf{r}$ is the true source location. The MSE and 
the error rate can reveal the differences in performance regarding precision 
and accuracy respectively, i.e. a consistent bias in the estimate against 
correct estimates. In the experiments, we have chosen $f_s = 48$~kHz, $R_a = 
0.035$~m, and $M = 6$ microphones. The angle-dependent system impulse response 
$\gamma_0(t,\varphi)$ is measured for 12 uniform angles and truncated to $N_v = 
50$ samples. The parameters of the image source candidate locations are set to 
$T = 78$ and $P = 2$ resulting in $\mathbf{h}$ having $936$ entries. 
The value for $\lambda$ parameter in Eq.~(\ref{eq:inverse_solution}) was 
determined empirically.
\vspace{-0.3cm}
\subsection{Results}
\subsubsection{Single Wall}
In the simulations, the wall is rotated $30^\circ$ around the system at a fixed 
distance of $0.5$~m to obtain $12$ wall-relative positions. This results in 
different DOAs and loudspeaker impulse responses. Experiments are made for 
each of the $12$ different DOA's and 50 different noise realizations per DOA. 
The results here reported are averaged over the whole set of $12\times50$ 
experiments. It is also assumed that the unknown image source is exactly 
located at one of the predefined points of candidate locations. 
We expect methods assuming an omnidirectional impulse response to 
introduce biases when the loudspeaker is directional \cite{tervo2012acoustic}. 
Indeed, Figures \ref{fig:res1a} and \ref{fig:res1b} show the results of 
locating 
a single wall versus the SNR. Methods not accounting for directionality have a 
fixed bias. This manifests itself in a higher error rate. GCC-PHAT 
specifically has here poor performance. This can be attributed to a TOA 
estimation that is 
not geometrically constrained by the microphone locations which makes the 
trilateration process more sensitive. This problem is significantly reduced in 
the proposed directivity-aware methods, namely CC-DAS and $\ell_1$-regularized 
LS.
\begin{table}[t]
\caption{Comparison of methods used in experimental setup. Methods with * denote 
proposed.}
\label{tab:methods}
\begin{tabular}{l|lll}
\multicolumn{1}{p{1.2cm}}{Method} & \multicolumn{1}{p{0.07\textwidth}}{LS 
Model} & \multicolumn{1}{p{0.07\textwidth}}{Deconvolution} & 
\multicolumn{1}{l}{Geometric loc.} \\ \hline
GCC-PHAT           & $\gamma_0(t)$         & GCC-PHAT                & LS 
Trilateration     \\
CC-DAS-OMNI    & $\gamma_0(t)$         & Cross Corr.       & Delay and Sum    
    \\
CC-DAS*       & $\gamma_0(t, \varphi)$ & Cross Corr.       & Delay and Sum  
      \\
$\ell_1$ reg. LS OMNI & $\gamma_0(t)$         & Sparse deconv.         & 
Plane wave resp.     \\
$\ell_1$ reg. LS*   & $\gamma_0(t, \varphi)$ & Sparse deconv.         & Plane 
wave resp.
\end{tabular}
\vspace{-0.5cm}
\end{table}
\subsubsection{Two Walls}
We place the system in a $90^\circ$-degree corner equidistant to the two walls 
forming it. As before, the system is also rotated to average over several 
realizations. We require each of the methods to provide two location 
estimates. GCC-PHAT is not considered here due to its poor performance in the 
single-wall case and the fact that TOAs would need to be sorted out to their
corresponding sources, which falls outside the scope of this work. In the 
two-walls case both reflections arrive very close in time at the microphones. 
Steered-response power methods usually have difficulty in this scenario 
\cite{tervo2012acoustic,torres2013room}. We expect sparse recovery methods to
be less affected by time smearing.
As can be seen in Fig.~\ref{fig:res1c} and \ref{fig:res1d}, methods not 
considering directivity are biased. When considering only MSE it seems that to 
be able to correctly resolve individual echoes is more relevant than 
the inclusion of a directivity model, which would explain why CC-DAS methods 
underperform here. The bias introduced, however, hinders performance with 
respect to accuracy. It can be observed that $\ell_1$-regularized LS 
consistently attains the best performance. This suggests that, by introducing a 
directivity model, sparse recovery methods are able to further reduce the 
uncertainty in echo detection.
\vspace{-0.2cm}
\section{Conclusions}
We present a method for robust estimation of reflecting boundaries that 
incorporates a loudspeaker directivity model. The times of arrival of echoes 
are 
estimated in a joint step where both microphone array geometry and
loudspeaker directivity are considered. The method is shown to 
decrease localization error and improve the accuracy compared to 
cross-correlation-based methods when more than one reflecting boundaries are 
present.
Future work can involve an extension of this method to multi-driver systems, 
and extending the model to account for room boundaries in three dimensions.
\vspace{-0.15cm}
\bibliographystyle{IEEEtran}
\bibliography{bibthesis}
\end{document}